# Simple Quantum Mechanical Phenomena and the Feynman Real Time Path Integral

A. Dullweber*, E.R. Hilf and E. Mendel§
*Physics Department, Carl von Ossietzky Universität Oldenburg*
*26111 Oldenburg, Germany*

The path integral formalism gives a very illustrative and intuitive understanding of quantum mechanics but due to its difficult sum over phases one usually prefers Schrödinger's approach. We will show that it is possible to calculate simple quantum phenomena by performing Feynman's sum over all paths staying entirely in real time. Once the propagator is obtained it is particularly easy to get the energy spectrum or the evolution of any wavefunction.

## I. INTRODUCTION

In 1942 R.P. Feynman presented the path integral formalism in configuration space as an alternative and very illustrative description of quantum mechanics [1,2]. Many applications in areas like quantum statistical mechanics or field theory were explored in the following years [3–7].

Feynman described the probabilistic character of quantum mechanics by the ability of a *quantum particle* to take virtually all possible paths between two points. The different paths were weighted by a phase factor, depending on the classical action of the path measured in units of $\hbar$. The sum over all paths (the path integration) of these phase factors results in the amplitude for the probability of propagation. In the classical limit of very large actions (in units of $\hbar$) the Hamilton principle of stationary action becomes apparent, as all other highly oscillating phase factors can interfere destructively.

But if we treat the genuine quantum mechanical case there arise severe mathematical difficulties due to the oscillating path phases. Many sophisticated methods have been developed to find the most important class of paths, e.g. stationary phase approximations (in semiclassical quantum mechanics) [13,8,9], or to damp the integrand (e.g. Euclidean (imaginary) time path integration) as used for finite temperature calculations [3,4,6].

In the present work we will explore the direct use of the basic concept of the sum over all paths in configuration space and *real* time. We will perform it without any weighting or damping change of the phase factors in our numerical calculations by iterating the 1-time-slice propagator. It will be seen that simple quantum systems such as the harmonic oscillator or the double well potential can be well described by this illustrative method.

The fact that one can perform the path integral directly in Minkowski space could turn out to be important for many applications like for real time tunnelling effects at finite temperature.

*New address: University of Cambridge, Department of Chemistry, Cambridge CB2 1EW, GB

§Dedicated to the memory of my beloved brother, Prof. Roberto Mendel

## II. CALCULATION OF THE REAL-TIME PROPAGATOR

### A. Discretization method

If we want to investigate the dynamics of a quantum system in configuration space we have to derive first the local representation of the time evolution operator $\hat{U}$. Then, by using

$$\Psi(x,t_\mathrm{b}) = <x|\Psi(t_\mathrm{b})> = \int \mathrm{d}x' <x|\hat{U}(t_\mathrm{b},t_\mathrm{a})|x'> \Psi(x',t_\mathrm{a}) \qquad (1)$$

we obtain for any given initial wavefunction $\Psi(x,t_\mathrm{a})$ the propagated one $\Psi(x,t_\mathrm{b})$ at a future time $t_\mathrm{b} > t_\mathrm{a}$.

The Feynman path integral,

$$<x_\mathrm{b}|\hat{U}(t_\mathrm{b},t_\mathrm{a})|x_\mathrm{a}> = \int_{(x_\mathrm{a},t_\mathrm{a})\quad(x_\mathrm{b},t_\mathrm{b})} \mathcal{D}x\; e^{\frac{i}{\hbar}\int_{t_\mathrm{a}}^{t_\mathrm{b}}\mathrm{d}t\mathcal{L}(x,\dot{x})}, \qquad (2)$$

enables us to get this propagator, $<x_\mathrm{b}|\hat{U}|x_\mathrm{a}>$, from the sum over all paths between $(x_\mathrm{a},t_\mathrm{a})$ and $(x_\mathrm{b},t_\mathrm{b})$, symbolised by $\mathcal{D}x$. The $\mathcal{L}$ denotes the classical Lagrangian of the system.

The fundamental composition law for the evolution operator suggests a discrete representation such as

$$\begin{aligned}<x_\mathrm{b}|\hat{U}(t_\mathrm{b},t_\mathrm{a})|x_\mathrm{a}> &= \int \mathrm{d}x(t_1)...\mathrm{d}x(t_{N-1}) \\ &\quad\times\prod_{n=1}^{N} <x(t_n)|\hat{U}(t_n,t_{n-1})|x(t_{n-1})> \\ &\simeq \lim_{N\to\infty} C^N \int \mathrm{d}x(t_1)...\mathrm{d}x(t_{N-1}) \\ &\quad\times\exp\left\{\frac{i}{\hbar}\sum_{n=1}^{N}\mathcal{S}(x(t_n),t_n,x(t_{n-1}),t_{n-1})\right\}, \end{aligned}\qquad (3)$$

with the equidistant discretization $(t_i), i = 0..N$, of the time interval $[t_\mathrm{a},t_\mathrm{b}]$ and ($t_\mathrm{a} \equiv t_0$, $t_\mathrm{b} \equiv t_N$, $x_\mathrm{a} \equiv x(t_0)$, $x_b \equiv x(t_N)$, $T := t_b - t_a$, $\Delta t := \frac{T}{N}$ and $C := (\frac{m}{2\pi i \hbar \Delta t})^{\frac{1}{2}}$. The evaluation of the action value $\mathcal{S}$ needs care in non trivial geometries [3–5] but also in numerical calculations with finite $N$. We use the *midpoint rule* which can be seen in Eq. 4 below.

Furthermore, we approximate the integrations over space by a finite equidistant discretization $(x_i)$, $i = 0..D$, on a spatial interval somewhat larger than the region where the wave functions could be of importance, $[x_0, x_D]$. We will treat only time independent potentials which causes all propagators to be functions of just time differences in our examples. In order to get the propagator we break it up into short time intervals. Thus we can use as a good approximation for the short time propagator the matrix $\mathcal{K}(\Delta t)$ defined by

$$\begin{aligned}\mathcal{K}_{ij}(\Delta t) &:= C \exp\left\{\frac{i\Delta t}{\hbar}\,\mathcal{L}\left(\frac{x_j + x_i}{2}, \frac{x_j - x_i}{\Delta t}\right)\right\} \\ &\approx\, < x_i(t_k + \Delta t)|\hat{U}(\Delta t)|x_j(t_k) >, \qquad (4)\end{aligned}$$

for arbitrary values of $x_i$ and $x_j$. We obtain then a universal $(D+1)$-dimensional, complex matrix $\mathcal{K}(\Delta t)$ independent of $t_k$.

Consequently, the approximated finite time propagator

$$\mathcal{G}_{ij}(T) \approx\, < x_i|\hat{U}(T)|x_j >$$

results by taking the product over short time propagators $\mathcal{K}(\Delta t)$ as

$$\mathcal{G}(T) := (\Delta x)^{N-1}\ \mathcal{K}^N(\Delta t)\ , \qquad (5)$$

with $\Delta x := \frac{x_D - x_0}{D}$, via simple matrix multiplications. The numerical effort can be substantially reduced by recursive matrix multiplication, needing therefore only $\ln(N/2)$ steps. Thus we can get now fairly precisely (with very small $\Delta t$) the probability amplitude of a *quantum particle* to go from $x_i$ to $x_j$ in time $T$.

Interesting to compute is the propagation of a discretized wavefunction $\Psi$ by inserting the approximation (5) into (1). With it we can measure the evolving expectation values,

$$\begin{aligned}&< O > (t) :=< \Psi|\hat{U}^+(t)\ \hat{O}\ \hat{U}(t)|\Psi > \\ &\approx \sum_{i,j,k,l} \Psi^*(x_i)\ \mathcal{G}^*_{ij}(t)\ \ < x_j|\hat{O}|x_k >\ \mathcal{G}_{kl}(t)\ \Psi(x_l), \qquad (6)\end{aligned}$$

for an arbitrary observable $\hat{O}$ in order to get physical information about its time evolution (e.g. the *particle* position or the conservation of energy).

### B. Energy Levels

One of the most important quantities to extract, for time independent systems, is the energy spectrum of the Hamiltonian $\hat{H}$. We can get this from the propagator by calculating its trace as a function of time. Taking a complete set of energy eigenstates $|n >$, with $n = 0, 1, ...$ and

$$\hat{H}|n >= E_n|n >\ , \qquad (7)$$

we can decompose

$$\begin{aligned}\mathrm{Tr}\ \hat{U}(t) &= \int \mathrm{d}x < x|e^{-\frac{i}{\hbar}\hat{H}t}|x > \\ &= \sum_n \int \mathrm{d}x < x|e^{-\frac{i}{\hbar}\hat{H}t}|n >< n|x > \\ &= \sum_n e^{-\frac{i}{\hbar}E_n t} \qquad (8)\end{aligned}$$

due to orthonormality of the wavefunctions. We obtain the energy levels after Fourier transformation[1] of the trace [11]. However, even in analytic calculations this yields serious trouble because of singularities and poles, and needs to be addressed with care.

The time development of the trace on the interval $[T, (N_T+1)T]$ can be performed recursively with the composition law,

$$\mathcal{G}(t_i) = \mathcal{G}(T)\ \mathcal{G}(t_{i-1})\ \Delta x, \qquad (9)$$

for the propagator matrix ($i = 1..N_T$ and $t_0 \equiv T$).

### C. Normalization

An estimate for the quality of the propagator can be obtained from the normalization condition, meaning, that the probability to go from one point to any possible point at a later time has to be 1, due to

$$\begin{aligned}1 &= \int \mathrm{d}x'\ \ < x'|x > \\ &= \int \mathrm{d}x'\mathrm{d}x''\ (< x''|\hat{U}|x' >)^* < x''|\hat{U}|x > . \qquad (10)\end{aligned}$$

In our discrete formulation we should get

$$\sum_{i,j=0}^{D} (\Delta x)^2\ \mathcal{G}^*_{ij}(t)\ \mathcal{G}_{ik}(t) = 1 \qquad (11)$$

for every starting point $x_k$.

In our examples we will see that this normalization is quite good in the needed spatial intervals. We will only use it to renormalize the long time propagator needed for the energy spectrum (Eq. 9) in order to compensate for numerical errors.

[1] We use the *Fast Fourier Transformation* [12].

## III. HARMONIC OSCILLATOR

The harmonic oscillator represents the simplest example where it is possible to confine the calculation to a finite position interval. Furthermore numerical results can be compared with analytic solutions.

In the numerical calculation we will use the scaled, one dimensional, classical Lagrangian $\mathcal{L}$,

$$\mathcal{L}(\dot{x}, x) = \frac{1}{2}\dot{x}^2 - \frac{1}{2}x^2. \tag{12}$$

with the resulting eigenfrequency $T_0 = 2\pi$. For a sufficient time resolution of an oscillation we will perform the propagator $\mathcal{G}_{ij}$ in intervals $T = \frac{T_0}{16}$ with 4 time slices (i.e. $N = 4$ and $\Delta t = \frac{T_0}{64}$ in Eq. 5, respectively).

First we investigate the time evolution for a real Gaussian wavefunction $\Psi_0(x)$,

$$\Psi_0(x) = \left(\frac{\alpha}{\pi}\right)^{\frac{1}{4}} e^{-\frac{\alpha}{2}(x - x_{\mathrm{Start}})^2}, \tag{13}$$

which should not change as a generalized coherent state its starting shape after a period. It can be regarded as a *quantum particle* at $< x >$ which oscillates with a period $T_0$. This will be confirmed up to an error of $< 10^{-6}$ % if we measure the position expectation values successively between the propagations (Fig. 1).

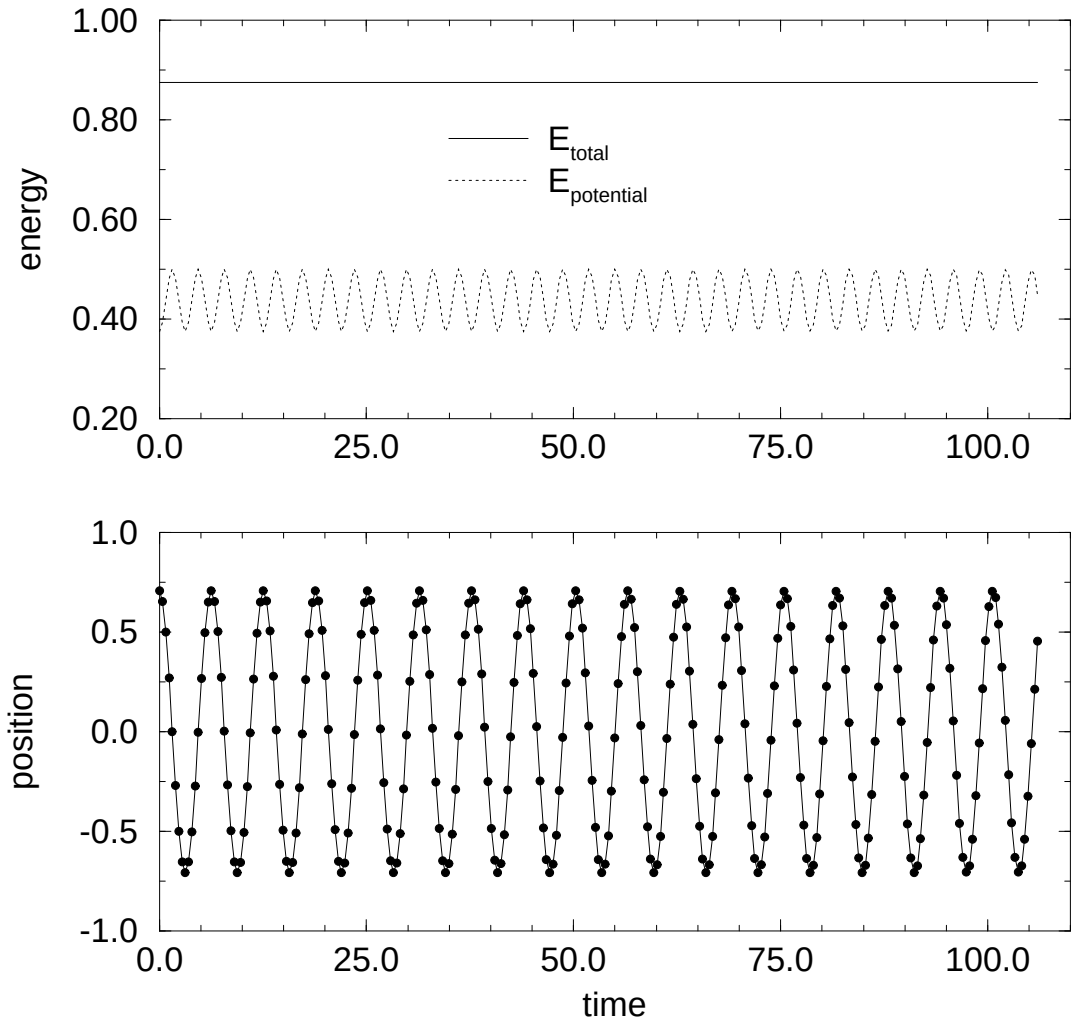

FIG. 1. Energy and position expectation values as a function of time for the harmonic oscillator ($\alpha = 2.0$).

Also is shown from measurements of the Hamiltonian that the conservation of energy is ensured up to $< 10^{-8}$ % per period. In addition, the virial theorem (e.g. in [10]) is well reproduced by estimating the time average of the potential energy.

On looking more carefully at Fig. 1 a difference between a classical and quantum *particle* can be extracted because the potential energy in quantum mechanics not necessarily possesses a maximum at the classical turning points. The ratio of the kinetic and potential energy depends on the shape of the wavefunction. This can be seen in Fig. 2, where the propagation of the probability density $|\Psi(x, nT)|^2$ for the above example (top) is shown. It is compared with a wavefunction which has a more classical-like time evolution of the energies (bottom). All numerical results are practically equal to the analytic ones.

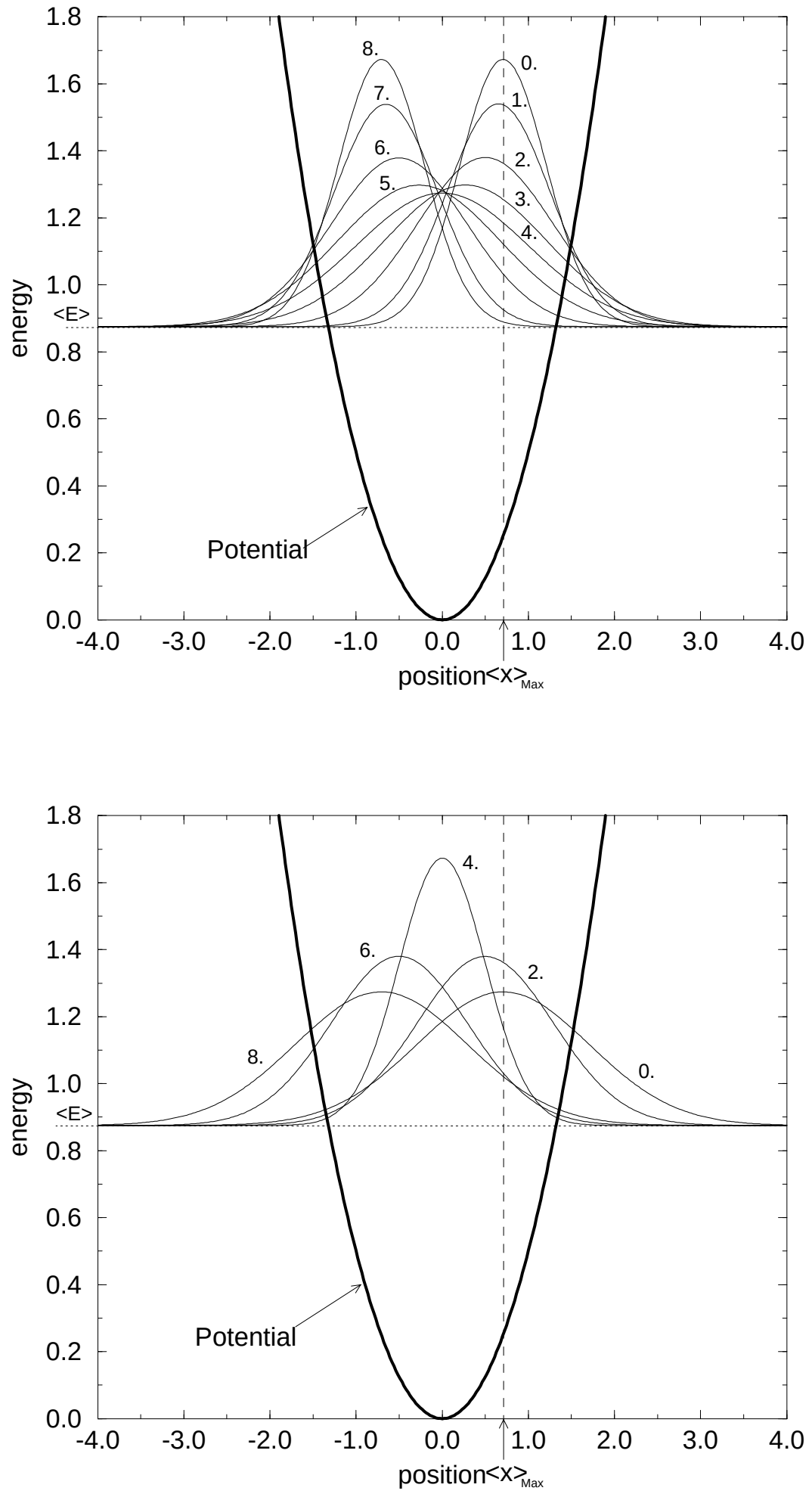

FIG. 2. Time development of the probability density of a Gaussian wavefunction $\Psi_n$ in the harmonic oscillator, with $\alpha = 2.0$ (top) and $\alpha = 0.5$ (bottom).

For the calculation of the energy spectrum from the propagator trace (Eq. 8) we constructed a kernel with $T = \frac{\pi}{7}$. The energy range of the spectrum is determined by this value only. It is not affected by estimating the trace evolution for a longer time period but this increases the accuracy of the spectrum. The peaks appear better localized and give more precise energy eigenvalues. The resulting graph from a discrete Fourier transformation of the trace in time is shown in Fig. 3. Here the evolution is calculated for a period of more than 70 harmonic oscillations.

The energy levels agree very well with the well known eigenvalues $E_n$,

$$E_n = n + \frac{1}{2}, \qquad n = 0, 1, \dots , \tag{14}$$

of the harmonic oscillator Hamiltonian. So in this case the main physical characteristics can be reproduced with high precision by our realization of the Feynman path integral *directly* in real time.

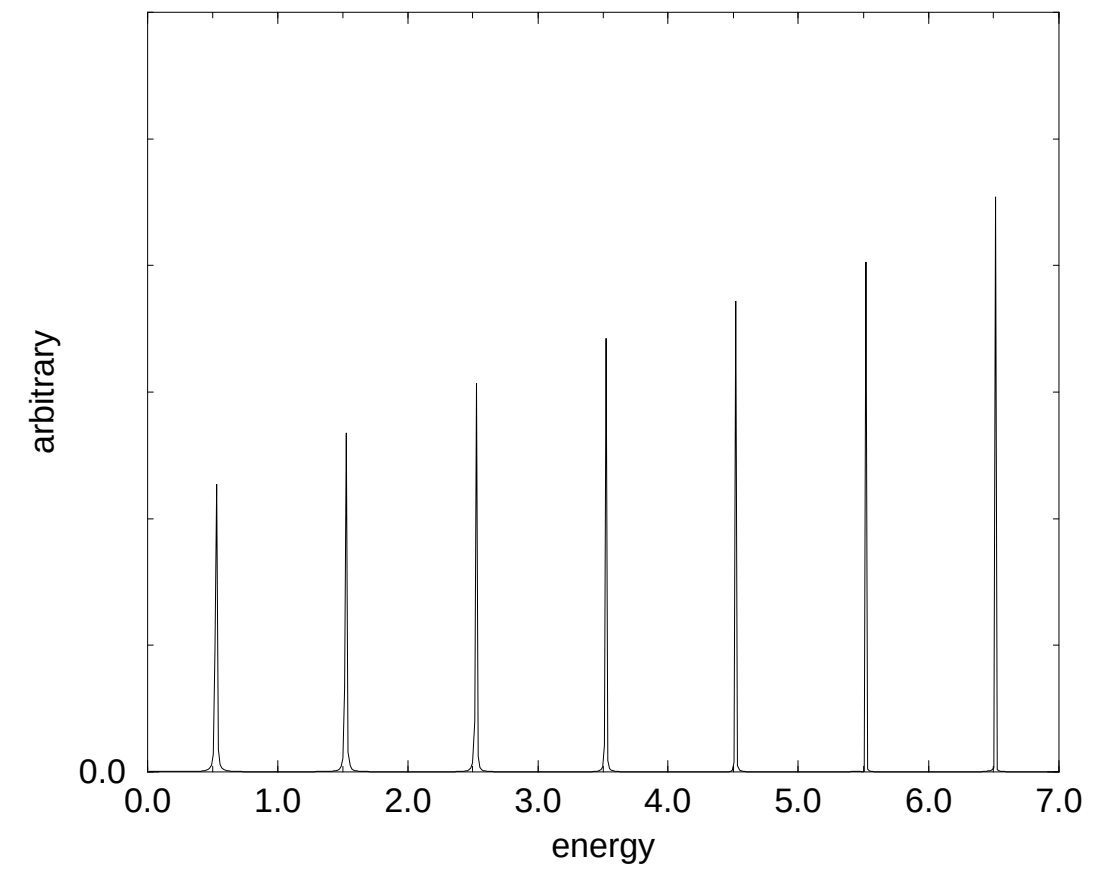

FIG. 3. Energy spectrum of the harmonic oscillator from Fourier transformation of the propagator trace ($N_T = 1023$).

Finally we show the propagator norm as a function of the starting position in Eq. 11 in Fig. 4.

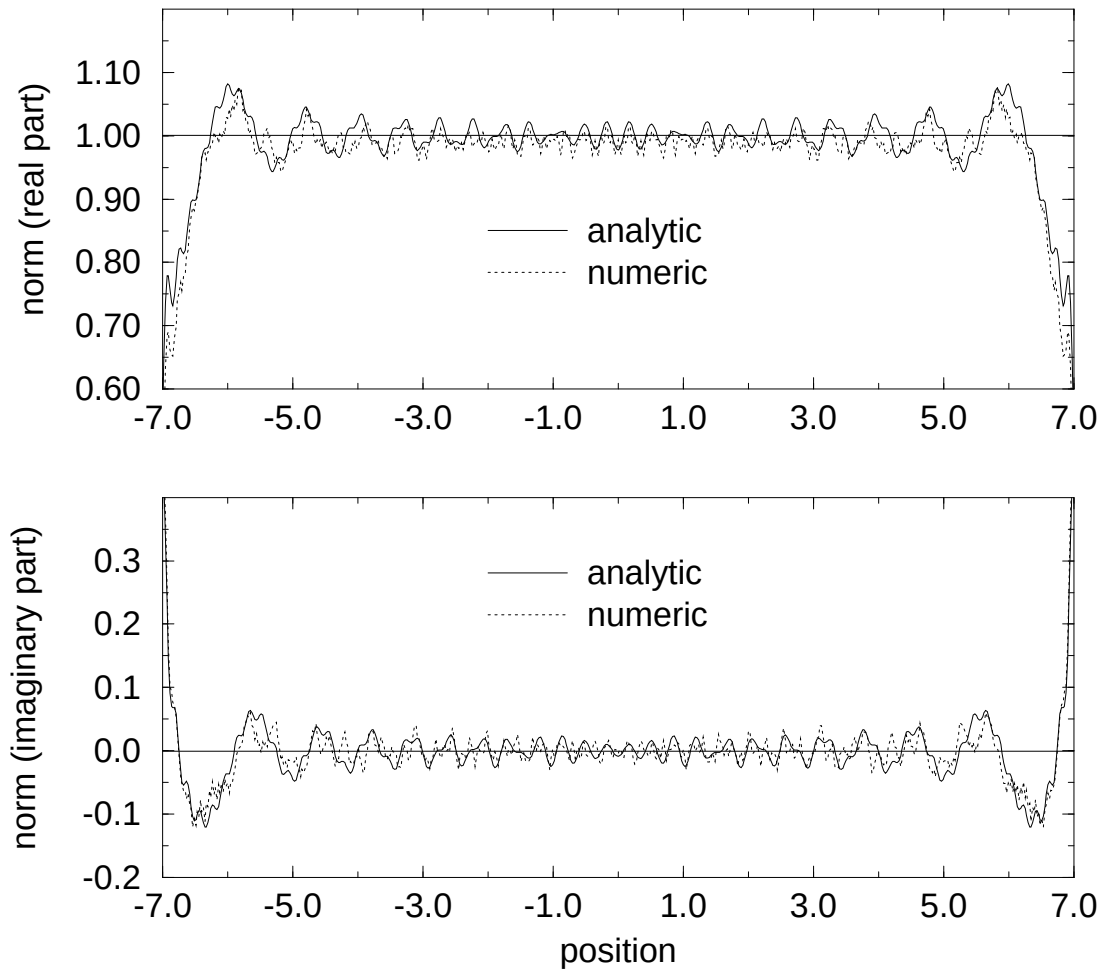

FIG. 4. Norm of a harmonic oscillator propagator compared to the exact solution on the interval $[-7, 7]$ for $T = \frac{T_0}{16}$.

Only after comparison with the exact continuous[2] solution (e.g. in [2,3,11]) on the same lattice we get an estimate for the rather high quality of the norm. It elucidates that the use of a finite spatial interval causes the main deviations from the perfect norm. Notice that we can always choose the interval wider than the relevant allowed region for a given potential thus avoiding this problem. In addition the norm provides a helpful convergence criterion for the choice of the spatial grid. In our examples 600 grid points are sufficient to minimize the influence of the discretization in position space.

## IV. DOUBLE WELL POTENTIAL

In a second example we examine the double well potential,

$$\mathcal{L}(\dot{x}, x) = \frac{1}{2}\dot{x}^2 - \alpha \left(x - x_{\mathrm{Min}}\right)^2 \left(x + x_{\mathrm{Min}}\right)^2 , \tag{15}$$

having two harmonic oscillator like minima. This *Mexican hat* potential is important in quite different fields (cf. e.g. [3]) but there exists no closed analytic solution for the propagator.

Here we want to show the time evolution of a tunnelling process as one of the most important phenomena in quantum mechanics. We construct an almost localized *particle* in one well, from a superposition of the lowest lying symmetric and antisymmetric energy eigenstates $\Psi_{\mathrm{S}}$ and $\Psi_{\mathrm{A}}$ of the total system. They can be approximated with two parametric testfunctions $T_{\mathrm{S/A}}$ ($T_{\mathrm{S}}$ '+' and $T_{\mathrm{A}}$ '−'),

$$T_{\mathrm{S/A}}(x) = \frac{1}{\alpha\sqrt{2\pi}} \left( e^{-\frac{(x-\beta)^2}{2\alpha^2}} \pm e^{-\frac{(x+\beta)^2}{2\alpha^2}} \right) , \tag{16}$$

by a simple variational method. The displaced Gaussians in Eq. 16 correspond to a harmonic approximation in the two wells. The resulting superposition is shown in Fig. 5.

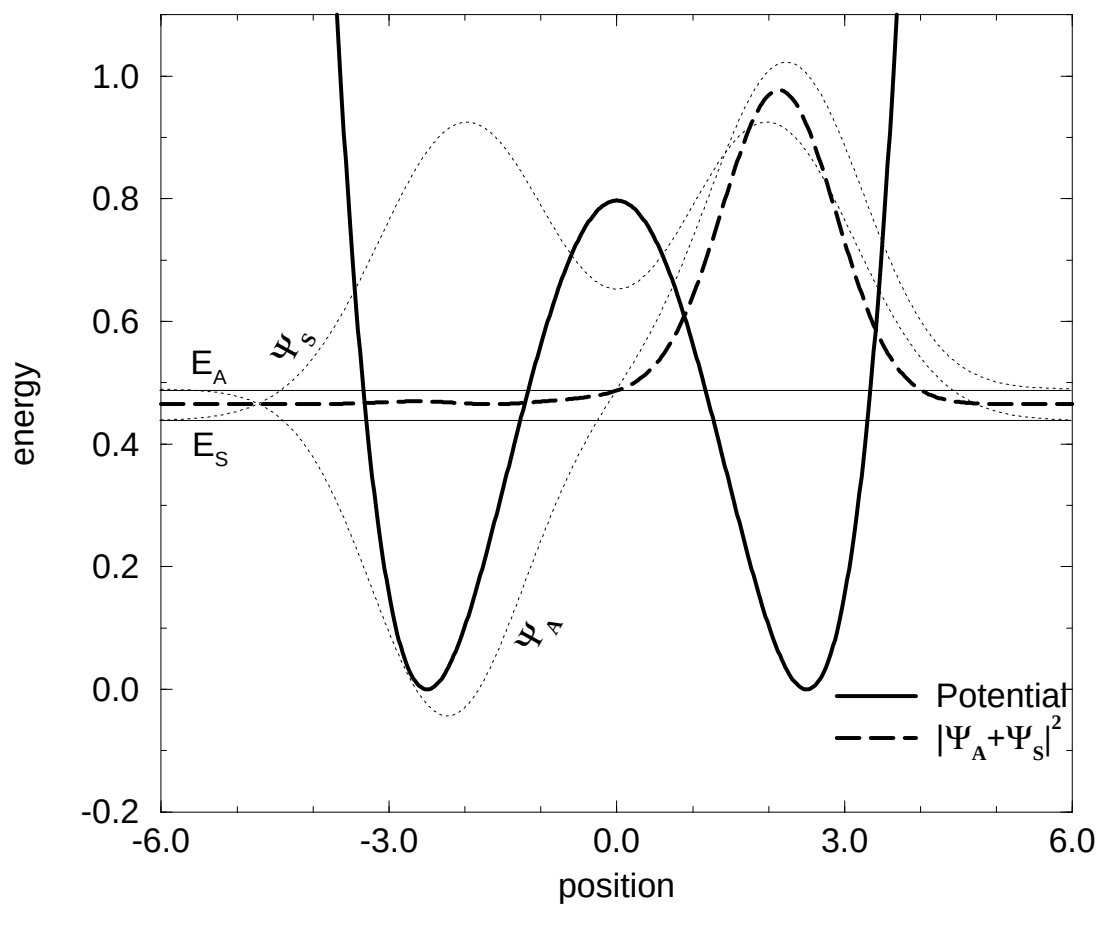

FIG. 5. Construction of a localized wavefunction on one side from an approximation for the lowest lying symmetric ($\Psi_{\mathrm{S}}$) and antisymmetric ($\Psi_{\mathrm{A}}$) energy states of the double well potential.

[2] In our example the discrete solution (cf. [3]) makes no significant difference.

Both energy levels are below the threshold in the potential centre and so a classical particle with the same energy could not leave one of the wells. But this is possible in the quantum case as can be shown using our calculated propagator (Fig. 6). Here every 55th time step of the squared wavefunction is plotted (i.e. $\Psi_n := \hat{U}^{55}(T)\Psi_0$).

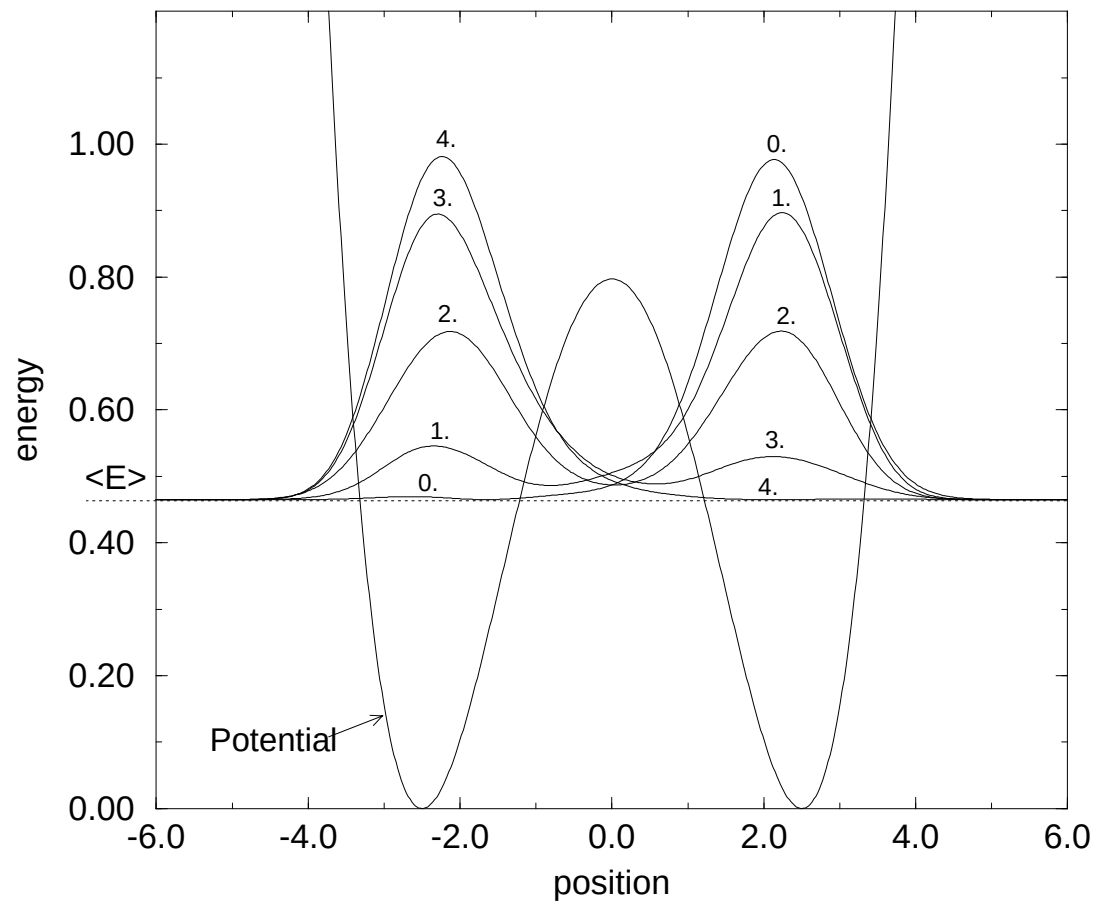

FIG. 6. Tunnelling of an initially ($\Psi_0$) on one side localized wavefunction $\Psi_n$ in a double well potential.

Decisive for the tunnelling rate is the size of the barrier which determines the also drawn energy difference of the two constituents $E_\mathrm{A} - E_\mathrm{S}$ (e.g. in the limit of very high barriers the energy levels become equal/degenerated). In our example the measurement of the position expectation value evolution (Fig. 7) results in a tunnelling time of $\tau = 54.4$. Furthermore the energy conservation is granted up to 1.2 % per tunnelling period.

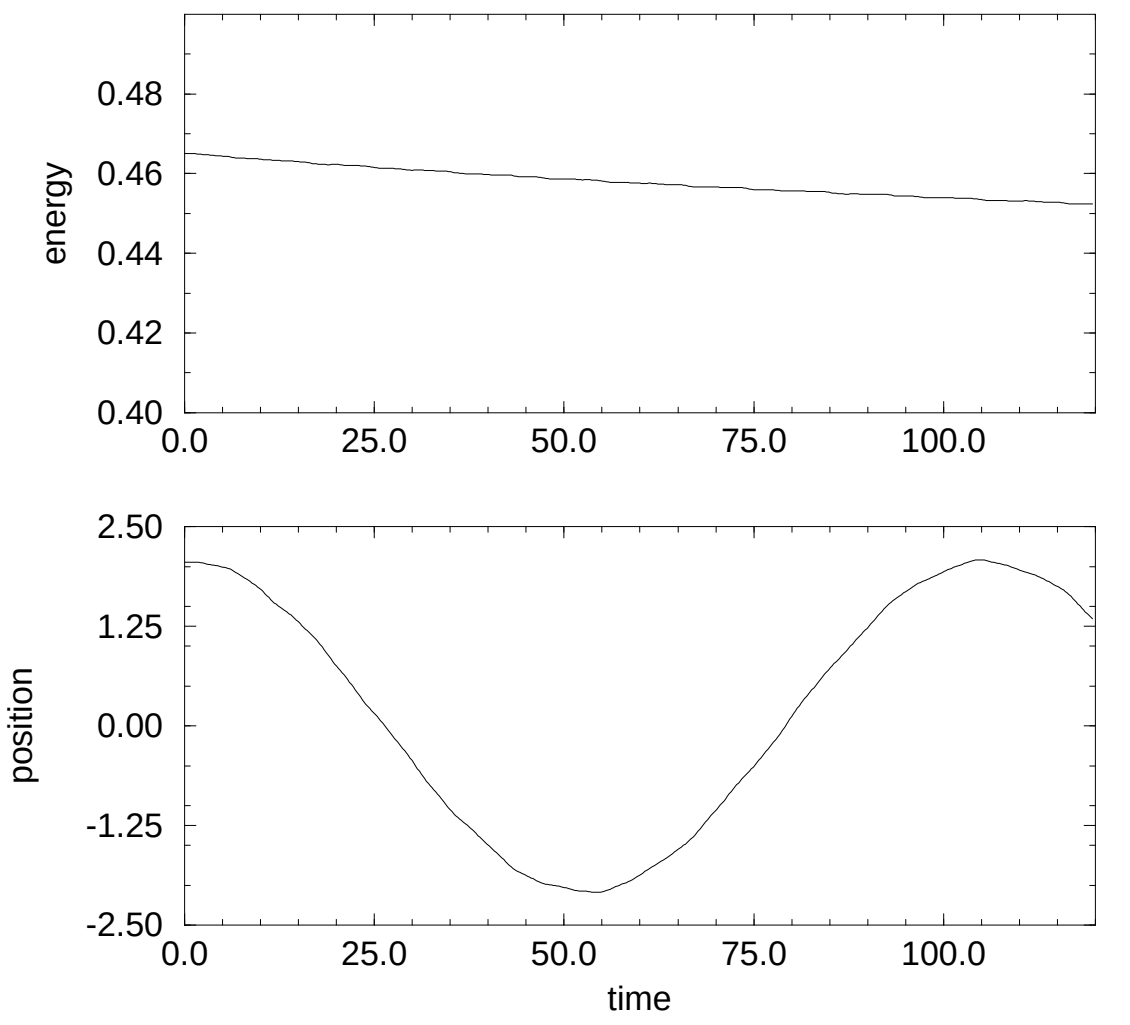

FIG. 7. Energy and position expectation value of a tunnelling wavefunction in a double well potential.

The tunnelling time can be checked using

$$
\begin{aligned}
\hat{U}(\tau)(\,|\Psi_\mathrm{S}> + |\Psi_\mathrm{A}>) \\
&= e^{i\tau\hat{H}}\,(|\Psi_\mathrm{S}> + |\Psi_\mathrm{A}>) \\
&= e^{i\tau E_\mathrm{S}}\left(|\Psi_S> + e^{i\tau(E_\mathrm{A}-E_\mathrm{S})}|\Psi_\mathrm{A}>\right), \qquad (17)
\end{aligned}
$$

assuming that we use the exact eigenstates. It is obvious that in a time $\tau_0 := \frac{\pi}{E_\mathrm{A}-E_\mathrm{S}}$ the sign in the superposition changes. This can be interpreted as a complete tunnelling to the other well. Precise measurements of the energy values $E_\mathrm{S}$ and $E_\mathrm{A}$ in our numerical example gives a tunnelling time of $\tau_0 = 61.4$. This 13 % deviation should result partly from the inaccurate approximation of the energy states $\Psi_{\mathrm{S/A}}$. This can be proved by a look at the energy spectrum of the double well (Fig. 8).

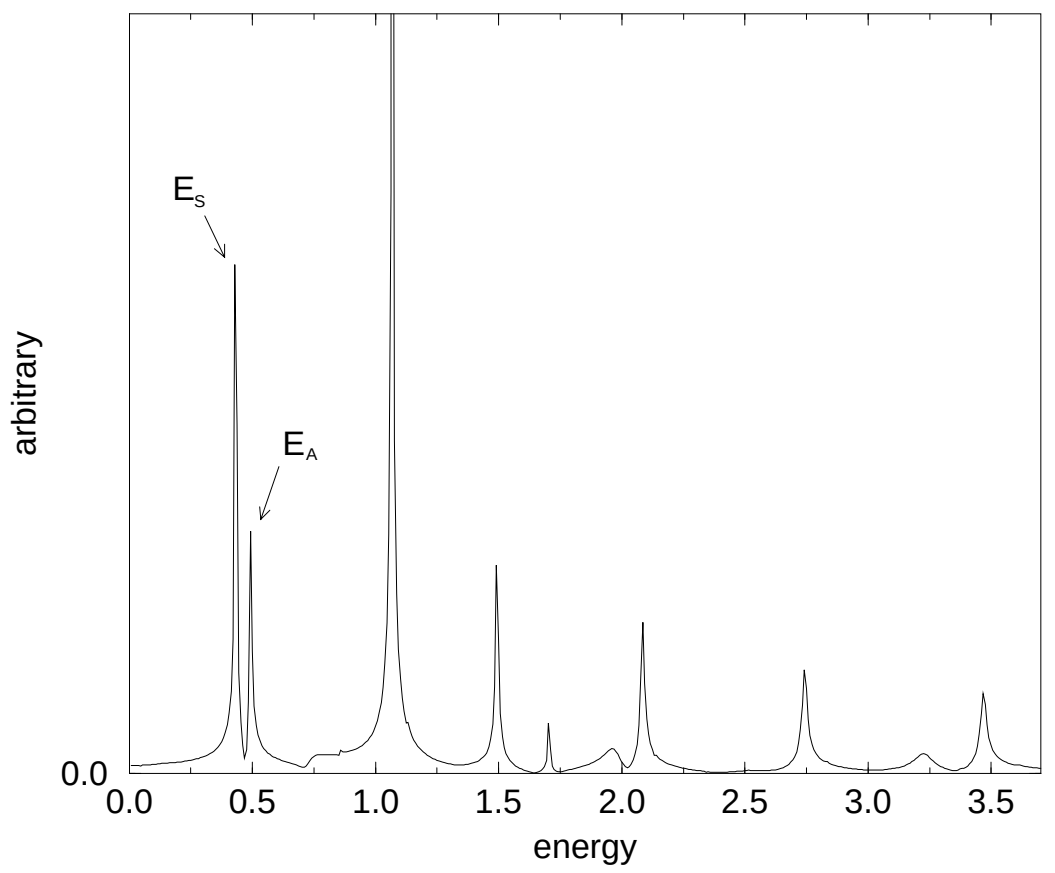

FIG. 8. Energy spectrum of the double well potential from Fourier transformation of the trace. The potential barrier is low enough to show the lowest lying eigenvalues $E_\mathrm{S}$ and $E_\mathrm{A}$ well separated($N_T = 2047$, $T = \frac{\pi}{8}$).

Easy to see is the broken degeneracy of the lowest lying eigenvalues $E_\mathrm{S}$ and $E_\mathrm{A}$ due to the finite barrier. The peaks occur at 0.433 and 0.494, respectively, close to our Gaussian superposition ($E_\mathrm{S} = 0.438$, $E_\mathrm{A} = 0.490$). However, the energy difference in the eigenspectrum results in a tunnelling period of $\tau_\mathrm{es} = 51.7$ which is only 5.0 % below our observed value $\tau = 54.4$ in Fig. 7. In addition the subsequent energy gaps are increasing as expected for all symmetric potentials which are of higher order than the harmonic one.

Finally the normalization shows also the correct shape in accordance with our experience from the harmonic oscillator (Fig. 9). In our calculations coarser or non-equidistant spatial matrices make it worse. But on the other hand a finer grid hardly improves the quality and the computational effort remains reasonable.

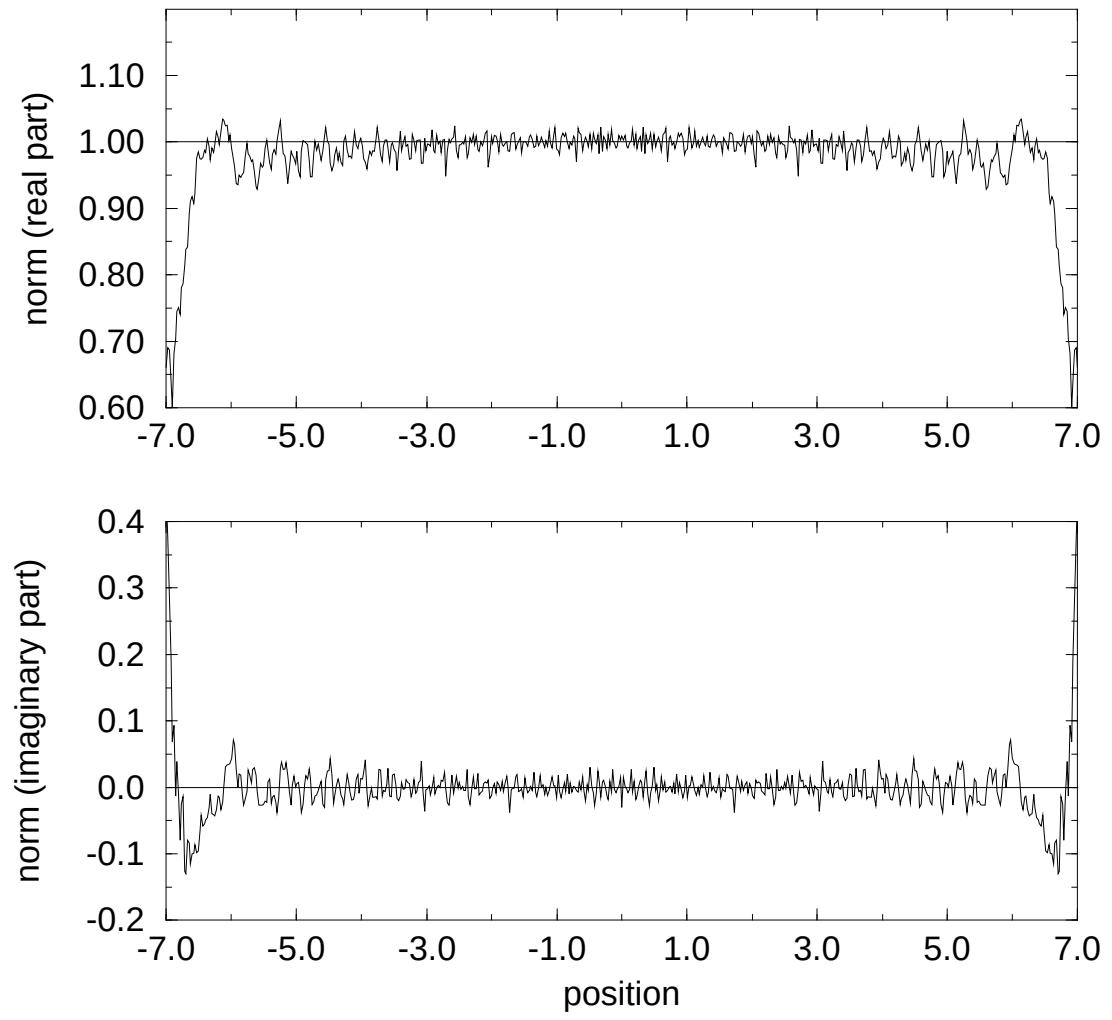

FIG. 9. Norm of a propagator in a double well potential on the interval $[-7, 7]$ ($T = \frac{\pi}{13}$).

Both examples show that it is possible to describe simple quantum phenomena with the illustrative Feynman real time path integral. It could be used not only as a didactic approach for the study of the elementary properties of quantum theories in cases where one needs the real time evolution, which it is hard to get by analytical continuation.

## V. ACKNOWLEDGEMENTS

A.D. is grateful to Holger Waalkens for stimulating discussions.